\documentclass[12pt]{article}

\usepackage{times}
\usepackage[utf8]{inputenc}
\usepackage[numbers]{natbib} 
\usepackage{hyperref} 

\usepackage[a4paper,margin=1in]{geometry}

\usepackage{amsmath, amssymb, amsthm, fullpage, enumitem}
\usepackage{authblk}

\usepackage{abstract}

\usepackage{titlesec}
\titlespacing*{\section}{0pt}{1.2em}{0.8em}
\titlespacing*{\subsection}{0pt}{1em}{0.6em}
\titleformat{\section}{\large\bfseries}{\thesection}{0.5em}{}{}

\title{\textbf{Enabling High-Frequency Trading with Near-Instant, Trustless Cross-Chain Transactions via Pre-Signing Adaptor Signatures}}
\author[1]{Ethan Francolla}
\affil{Cornell University\\
  \texttt{ef424@cornell.edu}}
\author[2]{Arnav Shah}
\affil{Cornell University\\
  \texttt{aks274@cornell.edu}}

\date{March 16, 2025\\[0.5em] v0.1}

\begin{document}
\maketitle

\begin{abstract}
Atomic swaps have been widely considered to be an ideal solution for cross-chain cryptocurrency transactions due to their trustless and decentralized nature. However, their adoption in practice has been strictly limited compared to centralized exchange order books because of long transaction times (anywhere from 20 to 60 minutes) prohibiting market makers from accurately pricing atomic swap spreads. For the decentralized finance ecosystem to expand and benefit all users, this would require accommodating market makers and high-frequency traders to reduce spreads and dramatically boost liquidity. This white paper will introduce a protocol for atomic swaps that eliminates the need for an intermediary currency or centralized trusted third party, reducing transaction times between Bitcoin and Ethereum swaps to approximately 15 seconds for a market maker, and could be reduced further with future Layer 2 solutions.
\end{abstract}

\vspace{1em}
\textbf{Keywords:} Atomic Swap \(\cdot\) Adaptor Signature \(\cdot\) Schnorr Signature \(\cdot\) Threshold Signature \(\cdot\) Taproot \(\cdot\) PTLC (Point Time-Locked Contract) \(\cdot\) HTLC (Hash Time-Locked Contract) \(\cdot\) Oracle \(\cdot\) HFT (High-Frequency Trading) \(\cdot\) Factory Smart Contract \(\cdot\) Bitcoin \(\cdot\) Ethereum

\section{Introduction}
Atomic swaps are protocols that enable trustless peer-to-peer exchanges of cryptocurrencies on different blockchains. They are known as "atomic" because of the binary nature of the outcome: a successful swap or a safe refund. Traditional atomic swap protocols have predominantly relied on Hash Time-Locked Contracts (HTLCs), which use explicit hash locks and time locks to ensure that funds can only be redeemed if a secret preimage is revealed before a timeout. While effective, HTLCs are constrained by long timeout periods, with successful swaps sometimes taking up to 60 minutes. More efficient solutions have been theorized, but have only been made possible with the addition of Bitcoin Taproot addresses and Schnorr signatures and their widespread adoption by mainstream wallets. In this white paper, we present a protocol that leverages Point Time-Locked Contracts (PTLCs) built on Bitcoin’s Taproot and Schnorr signature framework. Unlike HTLCs, PTLCs use adaptor signatures: a mechanism that “pre-signs” Bitcoin transactions with a hidden adaptor component. We were inspired by Gugger's work in establishing "unstoppable swaps" using PTLCs to trustlessly enable cross-chain exchanges \cite{Gugger2020} \cite{UnstoppableSwap_core}. The linearity of Schnorr signatures allows us to embed conditions off-chain (“scriptlessly”) enabling the final signature to be completed when an external condition is met such as the lock-up of ETH in a corresponding contract. This method not only reduces the overall waiting time (particularly for market makers), but also enhances privacy and flexibility, allowing for features such as collateral enforcement and reputation tracking.

\section{Background}

\subsection{Schnorr Signatures Overview}
Schnorr signatures are based on the difficulty of the discrete logarithm problem over an elliptic curve \cite{Schnorr1990}. 

\subsubsection{Preliminaries}
We define the following:
\begin{itemize}[label=\(\bullet\)]
    \item \(q\) be a large prime, the order of the elliptic curve group \(\mathbb{G}\).
    \item \(G \in \mathbb{G}\) be a generator of \(\mathbb{G}\).
    \item \(x \in \mathbb{Z}_q\) be the private key.
    \item \(P = xG\) be the corresponding public key.
    \item \(m \in \{0,1\}^*\) be the message to be signed.
    \item \(H: \{0,1\}^* \to \mathbb{Z}_q\) be a cryptographic hash function (e.g., SHA-256).
\end{itemize}

\subsubsection{Signature Generation}
The signing algorithm for a message \(m\) is as follows:
\begin{enumerate}
    \item \textbf{Nonce Generation:}  
          Choose a random nonce \(k \in \mathbb{Z}_q\) using a cryptographically secure random number generator, which will be kept secret. It is ephemeral, meaning a new random k is generated every time a signature is created. This ensures that even if the same message is signed twice, the signatures will differ. Compute the nonce point:
          \[R = kG.\]
          R acts as a public commitment to the random nonce k without revealing k itself.
    \item \textbf{Challenge Computation:}  
          Compute the challenge value \(e\) by hashing the concatenation of \(R\), \(P\), and \(m\):
          \[e = H(R \parallel P \parallel m),\]
          where \(\parallel\) denotes concatenation.
    \item \textbf{Signature Scalar Computation:}  
          Compute the signature scalar \(s\) as:
          \[s = k + ex \pmod{q}.\]
    \item \textbf{Final Signature:}  
          The signature is the pair
          \[\sigma = (R, s).\]
\end{enumerate}

\subsubsection{Signature Verification}

To verify a signature \(\sigma = (R, s)\) on a message \(m\) with respect to the public key \(P\), the verifier performs the following:

\begin{enumerate}
    \item Compute the challenge:
          \[e = H(R \parallel P \parallel m).\]
    \item Verify the equality:
          \[sG \stackrel{?}{=} R + eP.\]
\end{enumerate}

If the equation holds, the signature is valid; otherwise, it is rejected.

\subsection{Proof of Verification Equality}
We can be certain the signature verification equality is valid with the following proof, starting with the definition of the signature scalar:
\[s = k + ex \pmod{q}.\]
Multiply both sides by the generator \(G\):
\[
sG = (k + ex)G.
\]
Using the distributive property of scalar multiplication:
\[
sG = kG + (ex)G.
\]
Recall that:
\begin{itemize}
  \item \(kG = R\),
  \item \( (ex)G = e(xG) = eP \) (by associativity).
\end{itemize}
Thus, we obtain:
\[sG = R + eP.\]

\section{Our Proposal: Pre-Signing Adaptor Signatures}
Current Atomic swap protocols using HTLCs have the following properties:
\begin{itemize}[label=\(\bullet\)]
    \item \textbf{Long Wait Times and Inflexibility:} Due to differences in block times (e.g., Bitcoin’s $\sim$10 minute block time), HTLC-based swaps typically require long timeouts (often 60 minutes or more) to ensure both parties can properly lock up their funds. This inefficiency makes market making and high-frequency trading nearly impossible.
    \item \textbf{Script Complexity and Privacy Compromises:} HTLCs rely on explicit on-chain scripts that reveal the hash preimage on both blockchains, reducing privacy. A malicious observer can infer that the same hash appearing on both chains at approximately the same time indicates a swap between the two currencies.
\end{itemize}

\subsection{Advantages}
In our proposal, we leverage the advanced properties of Schnorr signatures and Bitcoin’s Taproot upgrade to implement \emph{pre-signed adaptor signatures}. This approach effectively transforms the traditional HTLC into a \emph{Point Time-Locked Contract (PTLC)} with the following key properties \cite{Gugger2020}, \cite{Tu2024}, \cite{You2024}:
\begin{enumerate}[label=\(\bullet\)]
    \item \textbf{Scriptless and Private Conditions:}
          Unlike HTLCs, which expose an explicit hash preimage on-chain, our PTLC leverages the linearity of Schnorr signatures to “lock” the final signature. In our scheme, Bob pre-signs his Bitcoin transaction with an adaptor signature that remains incomplete until an external unlocking value is provided. This means that the unlocking secret remains hidden until the swap condition (e.g., ETH lock-up) is verified, thus enhancing privacy.
          
    \item \textbf{Rapid Maker Finalization:} \\
          By requiring the taker (buyer) to initiate the swap and lock funds first, the seller (market maker) only finalizes their Bitcoin transaction once the external condition is met. Since the market maker generates the transaction, including all signatures and adaptor components, entirely off-chain, they can perform all necessary cryptographic computations without waiting for any block confirmations. Additionally, since the pre-signed transaction details also get sent to the taker off chain, the taker can quickly verify all parameters (including the commitment to the unlocking secret) are correct. The immediate feedback allows the taker to be confident in the swap's structure long before the transaction is broadcast on-chain. With our adaptor signature mechanism, the maker's finalization step, locking up funds in a smart contract so an oracle can release the unlocking value to their pre-signed signature, can occur in as little as 15 seconds (this time could be reduced even further in the future with Layer 2 implementations). This rapid execution is particularly beneficial for high-frequency trading and market making.
          
    \item \textbf{Atomic Cross-Chain Binding:} \\
          The final valid Bitcoin signature is computed as:
          \[
          s_{\text{final}} = s^* + \delta,
          \]
          where \( s^* \) is Bob's pre-signed (incomplete) signature and \( \delta \) is the adaptor component. The unlocking value \( \delta \) is only released by an external mechanism (e.g., a decentralized oracle or factory contract) after the buyer’s (Alice's) ETH has been securely locked. This binds the Bitcoin and Ethereum sides of the swap together in an atomic fashion—either both sides complete, or both are refunded.
          
    \item \textbf{Decentralized Verification via an Oracle:} \\
          To ensure that Bob cannot finalize his Bitcoin transaction without meeting the external condition of locking up his funds first, an external oracle (or a distributed threshold mechanism) is employed. This oracle, whose public key is \( O \), produces an unlocking signature that is incorporated into the Bitcoin transaction. The Bitcoin Taproot script verifies that the unlocking value, when combined with Bob’s pre-signed signature, yields a valid signature and that the unlocking value satisfies the pre-committed hash \( H(s_a) \). This external verification guarantees that the unlocking value originates from a trusted, decentralized source, not forged from Bob himself.
          
    \item \textbf{Efficient Contract Deployment and Economic Incentives:} \\
          We utilize factory contracts to deploy minimal swap instances at a low gas cost. By having the buyer initiate the swap instance, we ensure that the maker only incurs risk and completes his side when the buyer's funds are committed. Furthermore, mechanisms such as collateral posting and timeout penalties provide economic incentives for honest participation.
\end{enumerate}

This unique combination of several properties: scriptless enforcement, rapid finalization on the maker’s side, atomic binding across chains, and decentralized oracle verification is what distinguishes our approach from traditional HTLCs and enables efficient, high-frequency cross-chain atomic swaps.

\subsection{Preliminaries}

\begin{itemize}[label=\(\bullet\)]
    \item Let \(s_a \in \mathbb{Z}_q\) be an externally generated secret (also called the \emph{adaptor secret}).
    \item Define a function \(f: \mathbb{Z}_q \to \mathbb{Z}_q\) that produces the adaptor component. For simplicity, we set
          \[f(s_a) = s_a.\]
    \item The adaptor component is defined as:
          \[\delta = f(s_a) = s_a.\]
    \item A public commitment to the adaptor secret is given by:
          \[C = H(s_a),\]
          where \(H\) is a cryptographic hash function (e.g., SHA-256). This commitment \(C\) is shared with the counterparty.
    \item \textbf{Pre-Swap Conditions and Communication:} \\
          Prior to initiating the swap, the parties agree on the following:
          \begin{itemize}[label=--]
              \item \textbf{Swap Parameters:} Both parties agree on the asset amounts, timeouts, and other conditions necessary for the swap. Timeouts on both ends ensure if either party does nothing, they will receive their funds back eventually.
              \item \textbf{Role Assignment:} The protocol is designed such that the buyer (Alice) initiates the swap by locking her ETH, while the market maker (Bob) completes his side by finalizing his adaptor signature to send Bitcoin.
              \item \textbf{Proposal Exchange:} \\
                    Bob prepares a swap proposal message \(M\) containing:
                    \begin{itemize}[label=\(\circ\)]
                        \item The details of the swap (e.g., asset amounts, timeouts, etc.),
                        \item The public commitment \(C = H(s_a)\),
                        \item A reference to the pre-signed Bitcoin transaction using an adaptor signature.
                    \end{itemize}
              \item \textbf{Identity Verification:} \\
                    To ensure authenticity, Bob signs the message \(M\) with his reputable Ethereum private key. Let \(\sigma_{Bob}\) denote this signature. The message \(M\) together with \(\sigma_{Bob}\) is sent to Alice.
              \item \textbf{Verification by Alice:} \\
                    Alice verifies \(\sigma_{Bob}\) using Bob's known ETH public address. This step guarantees that the swap proposal indeed originates from the trusted party Bob. Although this protocol is trustless, this paper will later discuss why it helps if Alice knows Bob is a reputable market maker.
              \item \textbf{Swap Initiation:} \\
                    Upon successful verification, Alice initiates the swap by interacting with the factory contract, which creates a new swap smart contract instance listing her incorporating her Ethereum address as a beneficiary and the agreed parameters.
          \end{itemize}
\end{itemize}

\subsection{Adaptor Signature Generation}
Assume that:
\begin{itemize}
  \item Bob generates a Schnorr signature on message \( m \) with private key \( x \) as:
    \[
      s_B = k + e x \pmod{q},
    \]
    where:
    \begin{itemize}
      \item \( k \in \mathbb{Z}_q \) is a randomly chosen nonce,
      \item \( R = kG \) is the nonce point,
      \item \( P = xG \) is Bob's public key, and
      \item \( e = H(R \parallel P \parallel m) \) is the challenge.
    \end{itemize}
  \item Bob chooses an external secret \( s_a \in \mathbb{Z}_q \) and defines the adaptor component as
    \[
      \delta = f(s_a) = s_a.
    \]
  \item Bob computes his \emph{adaptor (partial) signature} as:
    \[
      s^* = s_B - \delta \pmod{q},
    \]
    so that the incomplete signature is
    \[
      \sigma^* = (R, s^*).
    \]
\end{itemize}

At this stage, the final signature is not yet valid. It can only be completed when the unlocking value \( \delta' \) is provided, i.e., 
\[s_{\text{final}} = s^* + \delta' \pmod{q}.\]
For the signature to be valid, we must have
\[s_{\text{final}} = s_B \quad \Longrightarrow \quad \delta' = \delta.\]

\subsection{Oracle's Role in Unlocking}

An external oracle monitors an Ethereum smart contract (or another external condition) and releases the unlocking value \( \delta' \) only when that condition is met \cite{Chainlink2.0}. However, to prevent Bob from simply using his own knowledge of \( s_a \) (and thus \( \delta \)) to finalize the signature, the oracle is required to produce a cryptographic signature on a predetermined message that includes the commitment information. Let:
\begin{itemize}
  \item \( m_O \) be a message that encapsulates swap-specific parameters (e.g., swap ID, \( H(s_a) \), and other relevant data).
  \item The oracle has a private key \( o \) and corresponding public key \( O \).
  \item The oracle produces a signature \( \sigma_O = (R_O, s_O) \) on \( m_O \) such that:
    \[s_O G = R_O + e_O O,\]
    where
    \[e_O = H(R_O \parallel O \parallel m_O).\]
\end{itemize}
The unlocking value \( \delta' \) is released together with the oracle’s signature. Thus, Bob can only finalize his Bitcoin transaction if he includes both:
\begin{enumerate}
  \item The unlocking value \( \delta' \) (which should equal \( s_a \)), and
  \item The oracle signature \( \sigma_O \) that verifies under \( O \) for message \( m_O \).
\end{enumerate}

\subsection{Verification in the Taproot Script}

The Bitcoin Taproot spending condition is designed to check both the validity of the final signature and the authenticity of the oracle's unlocking signature. The verification procedure includes the following steps:

\begin{enumerate}
  \item \textbf{Final Signature Verification:} \\
        The final signature is computed as:
        \[s_{\text{final}} = s^* + \delta' \pmod{q}.\]
        The standard Schnorr verification equation requires:
        \[s_{\text{final}} G = R + eP,\]
        where \( e = H(R \parallel P \parallel m) \). This ensures that the signature is valid if and only if:
        \[s^* + \delta' = s_B \quad (\text{with } s_B = k + ex \pmod{q}).\]
  \item \textbf{Unlocking Value Commitment Check:} \\
        The script additionally checks that the unlocking value \( \delta' \) satisfies:
        \[H(\delta') = C,\]
        where \( C \) is the pre-committed hash \( C = H(s_a) \). This guarantees that \( \delta' \) is exactly the secret Bob committed to, i.e. \( \delta' = s_a \).
  \item \textbf{Oracle Signature Verification:} \\
        The Taproot script further requires that the unlocking value is accompanied by a valid oracle signature \( \sigma_O \). As a reminder, \( \sigma_O = (R_O, s_O)\). Using the known public key \( O \), the script verifies that:
        \[s_O G \stackrel{?}{=} R_O + e_O O,\]
        where \( e_O = H(R_O \parallel O \parallel m_O) \). Since the oracle's private key is never exposed, Bob cannot forge \( \sigma_O \). This verification ensures that the unlocking value \( \delta' \) was released by the oracle and not computed solely by Bob. 
        The final signature is valid if:
        \[s_{\text{final}} = s^* + \delta' = s_B,\]
        which implies that the unlocking value must satisfy:
        \[\delta' = \delta = s_a,\]
        and the oracle signature verifies under \( O \) for \( m_O \). Even though Bob knows \( s_a \), he cannot produce a valid oracle signature \( \sigma_O \) to complete the transaction without access to the oracle's private key. Therefore, the Taproot script’s combined checks ensure that the final Bitcoin signature is only valid if the unlocking value is genuinely provided by the oracle, which requires Bob to have locked up the required funds as a precondition.
\end{enumerate}

\subsection{Adaptor Signature Finalization}

When the external condition is met (for example, an oracle confirms that the required ETH is locked), the unlocking value \( s_a \) is released.

\begin{enumerate}
    \item Bob (or an external observer) retrieves \( s_a \) from the unlocking mechanism.
    \item The final signature is then computed as:
          \[s_{\text{full}} = s^* + s_a \pmod{q}.\]
    \item The completed signature is:
          \[\sigma = (R, s_{\text{full}}),\]
          which can be verified using the standard Schnorr verification equation:
          \[s_{\text{full}} G \stackrel{?}{=} R + eP.\]
\end{enumerate}

\subsection{Secret Extraction and Resistance to Malicious Third Parties}
One of the important properties of adaptor signatures is that, once the final signature 
\(\sigma = (R, s_{\text{final}})\) is published, anyone who has seen the corresponding 
adaptor (partial) signature \(\sigma^* = (R, s^*)\) can extract the unlocking secret by computing
\[s_a = s_{\text{final}} - s^* \pmod{q}\] \cite{Hoenisch2021}.
This extraction is enabled by the linearity of Schnorr signatures. 
\subsubsection{How the Smart Contract is Protected}
Even if a malicious third party (Eve) obtained \( s_a \), this does not compromise the swap. The Ethereum swap contract was deployed with Alice's beneficiary address as an immutable parameter. This contract governs the ETH side of the swap such that, upon successful execution, the ETH funds are released \emph{only} to Alice's address. Thus, even if Eve learns \( s_a \) and used it to force the smart contract to release its funds, it would only deposit the ETH to Alice's predetermined address, which is the desired outcome. 
\subsubsection{How the Taproot Output is Protected}
Similarly, the Taproot output enforces that funds can only be spent by a signature that is irrevocably bound to Bob's intended receiving address. Below is a description of this signature.
\begin{itemize}
    \item Let \( x \in \mathbb{Z}_q \) be Bob's private key,
    \item Let $G$ be the standard generator of the elliptic curve group used by Bitcoin (secp256k1),
    \item let \( P = xG \) be Bob's corresponding base public key.
\end{itemize}
To tie the spending of funds to specific parameters of a cross-chain swap, we incorporate the swap's metadata:
\[m_{\text{swap}} = \Bigl(\mathit{assetAmounts}, \,\mathit{timeouts}, \,C, \,\dots\Bigr),\]
where $C = H(s_a)$ is the hash commitment to the external secret $s_a$, and any other relevant fields necessary to identify this particular swap instance. Following best practices outlined in BIP-341\cite{bip341_github}, we use a \emph{tagged hash} to ensure domain separation. We define $t$ as an integer in $\mathbb{Z}_q$:
\[t = \text{TagHash}\Bigl(\text{``TapTweak''},\; (P \parallel m_{\text{swap}})\Bigr) \pmod{q},\]
where $\text{TagHash}(\cdot,\cdot)$ is a cryptographic hash function. 
It has a domain-separating tag ``TapTweak'' and \(\parallel\) denotes byte concatenation.
Let $T$ be the corresponding curve point for the integer $t$:
\[T = t\,G.\]
The \emph{tweaked} public key is then defined as:
\[P_{\text{tweak}} = P + T.\]
This ensures that the public key controlling the UTXO is no longer just $P$, but $P$ \emph{tweaked} by the value $T$ derived from the swap-specific data. Correspondingly, Bob's \emph{tweaked} private key is 
\[x_{\text{tweak}} = x + t \mod q,\]
satisfying
\[P_{\text{tweak}} = x_{\text{tweak}}\,G\quad\Longleftrightarrow\quad
P + T = (x + t)\,G.\]
This means only an entity knowing $x_{\text{tweak}}$ can produce valid signatures that spend the UTXO locked under $P_{\text{tweak}}$. When Bob (or a party holding $x_{\text{tweak}}$) spends the BTC, the Taproot script checks:
\[sG \stackrel{?}{=} R \;+\; e\,P_{\text{tweak}},\]
where $e = H(R \parallel P_{\text{tweak}} \parallel m)$ is the standard Schnorr challenge over $(R, P_{\text{tweak}}, m)$. This confirms that the spender holds the discrete log of $P_{\text{tweak}}$ (i.e., knows $x_{\text{tweak}}$) and thus is the rightful recipient of the UTXO, respecting all the swap conditions encoded in $m_{\text{swap}}$. If an attacker, Eve, tries to bypass the oracle or exploit \( s_a, \) they still cannot produce a valid Schnorr signature that still verifies under $P_{\text{tweak}}$ because they do not know $x_{\text{tweak}}$. Using $P_{\text{tweak}}$ as opposed to $P$ provides the following key benefits:
\begin{itemize}[label=\(\bullet\)]
    \item \textbf{Binding Funds to the Swap Parameters:}
    Since $t$ is derived from $m_{\text{swap}}$, altering the swap details (e.g., timeouts, amounts, or the commitment $C$) changes the tweak. This enforces that the final spend condition is tied \emph{exactly} to the agreed-upon parameters. This makes it impossible for anyone to alter the swap parameters without invalidating the spending key.
    \item \textbf{Resistance to Secret Leakage:}
    Even if the external secret $s_a$ (i.e., the adaptor secret) is leaked, an attacker cannot spend Bob's UTXO. They would still require knowledge of Bob's private key $x$ to calculate $x_{\text{tweak}}$ and produce a valid signature under $P_{\text{tweak}}$.
    \item \textbf{Scriptless, Private Enforcement:}
    Taproot allows script-level logic to be encoded within the public key itself, rather than an on-chain \texttt{OP\_IF} / \texttt{OP\_CHECKSIG}. Observers can see only $P_{\text{tweak}}$ and a valid Schnorr signature, revealing no explicit HTLC-like script or $s_a$. They cannot see if there was a swap condition or partial signature behind it.
\end{itemize}

\subsection{Reversed Swap Process While Maintaining Buyer-First Initiation}
In this reversed process, the roles are arranged such that:
\begin{itemize}[label=\(\bullet\)]
    \item \textbf{Alice (Buyer):} Initiates the swap by locking her ETH in a smart contract. The ETH contract is configured with Bob as the beneficiary, so that once the swap is executed, the ETH is paid out to Bob.
    \item \textbf{Bob (Market Maker):} Creates a pre-signed Bitcoin transaction that sends BTC to Alice. This transaction is constructed using an adaptor signature that remains incomplete until an external Oracle confirms that Alice has locked her ETH.
\end{itemize}

The key steps of the process are as follows:

\subsubsection{Bob Prepares the Bitcoin Transaction}
\begin{itemize}[label=\(\bullet\)]
    \item Bob selects a random nonce \(k \in \mathbb{Z}_q\) and computes the nonce point \(R = kG\).
    \item He computes the full Schnorr signature:
    \[s_B = k + ex \pmod{q},\]
    where:
    \begin{itemize}
        \item \(x\) is Bob's private key,
        \item \(P = xG\) is Bob's public key, and
        \item \(e = H(R \parallel P \parallel m)\) is the challenge, with \(m\) representing the swap details.
    \end{itemize}
    \item Bob then chooses an external unlocking secret \(s_a \in \mathbb{Z}_q\) and sets the adaptor component as:
    \[\delta = s_a.\]
    \item He computes his adaptor (partial) signature as:
    \[s^* = s_B - \delta \pmod{q}.\]
    \item Bob’s pre-signed Bitcoin transaction uses the adaptor signature \(\sigma^* = (R, s^*)\) and is constructed so that the spending condition in the Taproot output directs BTC to \emph{Alice}’s Bitcoin address.
\end{itemize}

\subsubsection{Bob Sends the Swap Proposal to Alice}
\begin{itemize}[label=\(\bullet\)]
    \item Bob packages the following information into a swap proposal message \(M\):
    \begin{itemize}
        \item The adaptor signature \(\sigma^* = (R, s^*)\),
        \item The public commitment \(C = H(s_a)\), and
        \item The swap parameters (asset amounts, timeouts, etc.).
    \end{itemize}
    \item Bob signs \(M\) with his reputable ETH private key to prove his identity and sends the signed message to Alice.
\end{itemize}

\subsubsection{Alice Initiates the Swap on Ethereum}
\begin{itemize}[label=\(\bullet\)]
    \item Upon verifying Bob's signed swap proposal (using off-chain tools such as Etherscan), Alice interacts with a factory contract to create a new swap instance.
    \item This swap instance is configured with:
          \begin{itemize}
              \item Bob's ETH beneficiary address (immutable),
              \item The commitment \(C = H(s_a)\), and
              \item The agreed swap parameters.
          \end{itemize}
    \item Alice then locks her ETH in the smart contract, thereby activating the swap.
\end{itemize}

\subsubsection{Oracle Release and Finalization of the Bitcoin Signature}
\begin{itemize}[label=\(\bullet\)]
    \item Once the ETH lock-up is confirmed, an external Oracle (or the mechanism integrated within the factory contract) verifies that the condition is met and releases the unlocking secret.
    \item Alice uses her Bitcoin wallet to redeem the BTC from the Taproot output. In doing so, she provides the unlocking value that completes the adaptor signature.
    \item The final signature is computed off-chain as:
    \[s_{\text{final}} = s^* + s_a \pmod{q}.\]
    Since \( s^* = s_B - s_a \), we have:
    \[s_{\text{final}} = s_B.\]
    \item The Bitcoin Taproot script enforces that the final signature \(\sigma = (R, s_{\text{final}})\) must verify as:
    \[s_{\text{final}}G = R + eP_{\text{tweak}},\]
    where \( P_{\text{tweak}} \) is a tweaked public key bound to Alice's receiving address (set during contract creation). This ensures that the BTC can only be redeemed by the intended conditions.
    \item The act of redeeming the BTC reveals \( s_a \) (since any observer can compute \( s_a = s_{\text{final}} - s^* \pmod{q} \)).
\end{itemize}

\subsubsection{Bob Claims the Locked ETH}
\begin{itemize}[label=\(\bullet\)]
    \item After Alice redeems the Bitcoin transaction, the unlocking secret \( s_a \) becomes public.
    \item Bob, monitoring the Bitcoin blockchain, extracts \( s_a \) from the final signature.
    \item Bob then uses \( s_a \) to trigger the payout from the Ethereum smart contract. Since the contract is configured with Bob as the beneficiary, the locked ETH is transferred to him.
\end{itemize}

\section{Discussion}
The proposed pre-signed adaptor signature scheme offers transformative potential for atomic swaps, enabling high-frequency trading (HFT) in DeFi without reliance on centralized order books. This innovative approach addresses several key limitations of traditional HTLC-based swaps, thereby increasing liquidity, enhancing security, and simplifying user experiences. Below, we detail how our method achieves these benefits:

\subsection{Enabling High-Frequency Trading and Enhanced Liquidity}

Traditional atomic swaps relying on HTLCs often require long timeout periods due to Bitcoin’s block times (typically 10 minutes per block), resulting in overall swap times that may exceed 60 minutes \cite{UnstoppableSwap_core}. This latency hinders the participation of high-frequency traders (HFTs) and market makers, who require rapid execution to minimize risk and capture arbitrage opportunities. In contrast, our scheme decouples the on-chain confirmation delay from the market maker’s (Bob's) critical operations:
\begin{itemize}[label=\(\bullet\)]
    \item \textbf{Buyer-First Initiation:} The swap is initiated by the buyer (Alice), who creates the ETH smart contract and sends her BTC to the taproot address created by Bob. The contract is created through a factory contract and takes only around 15 seconds on Ethereum mainnet. This is further reduced to 1--2 seconds on a future Layer~2 implementation.
    \item \textbf{Rapid Maker Finalization:} Once Alice's commitment is confirmed with the ETH contract and she's locked her BTC, Bob finalizes his Bitcoin transaction using pre-signed adaptor signatures after he locks up his ETH. The finalization (i.e. combining his partial signature with the unlocking secret from an external oracle) after locking funds in the smart contract can occur in as little as 15 seconds (lock-up time). Thus, despite having to wait around 10 minutes for Alice's BTC to confirm, from the market maker's perspective, no action is required during and before confirmation. Furthermore, the critical actions for executing the swap are nearly instantaneous.
\end{itemize}

The overall effect is a system that can provide liquidity on a trustless, decentralized basis without relying on centralized exchange order books. This increased liquidity benefits all participants, as it reduces slippage and makes it easier to swap crypto assets quickly and securely.

\subsection{Broader Consequences for Crypto Trading}

Our proposed protocol not only enhances speed and liquidity but also brings several other advantages:
\begin{itemize}[label=\(\bullet\)]
    \item \textbf{Security and Atomicity:} The use of adaptor signatures ensures that the Bitcoin transaction remains incomplete until the external condition (e.g., ETH lock-up) is met. This atomicity guarantees that either both sides of the swap execute or neither do, significantly reducing counterparty risk.
    \item \textbf{Enhanced Privacy:} Unlike traditional HTLCs, which reveal the hash preimage explicitly on-chain, our "scriptless" approach conceals the unlocking secret until finalization. This improves privacy for both parties \cite{Gugger2020}, \cite{You2024}.
    \item \textbf{Ease of Use:} Although our scheme employs advanced cryptographic techniques, the end-user experience remains simple. The buyer only interacts with a user-friendly interface to create the smart contract and lock BTC, while the maker’s operations can be easily automated.
\end{itemize}

\subsection{Considerations for Oracle Implementation}

A crucial component of our system is the external oracle that releases the unlocking secret. While the oracle may require integration with a third token or rely on a decentralized network of nodes (such as Chainlink, DIA, or another solution), the key point is that there exist multiple robust systems for this purpose. The oracle's role is to verify that the ETH smart contract conditions are met and to release the unlocking value accordingly. This flexibility means:
\begin{itemize}[label=\(\bullet\)]
    \item The protocol is not dependent on a single centralized entity, reducing risk.
    \item Multiple decentralized oracle solutions are available, and future innovation may further enhance efficiency and lower costs.
    \item Although one could design a custom token or ecosystem around the oracle mechanism, existing solutions already provide the necessary functionality and security, ensuring that our system is robust and not merely a tool for opportunistic schemes.
\end{itemize}

\subsection{Potential Risks for Both Parties}

While our proposed pre-signed adaptor signature scheme is designed to minimize risk and favor market makers, there remain some potential risks that must be considered for both parties. Below we discuss these risks in detail and explain how our design mitigates them.

\subsubsection{Risk Profile for the Market Maker (Bob)}

\begin{itemize}[label=\(\bullet\)]
    \item \textbf{Deferred Action and Zero Initial Risk:} \\
          Bob, acting as the market maker, does not need to lock up any funds until after Alice initiates the swap by deploying the Ethereum smart contract and locking her Bitcoin. This \emph{buyer-first} initiation minimizes Bob’s exposure since he only acts when the swap is active. Consequently, Bob faces minimal risk since his capital is not committed until the swap is underway. This can give the market maker confidence their capital will only be deployed towards swaps that can be profitable to them and not "waste" capital on swaps that will time out.
    
    \item \textbf{Market Volatility:} \\
          Once the external condition is met (e.g., Bob locking his ETH in roughly 15 seconds), Bob redeems his Bitcoin transaction almost immediately. This rapid execution minimizes his exposure to market fluctuations, further reducing his risk. Although this time is significantly faster than current atomic swap times and allows Bob to price tighter spreads due to less prolonged exposure to volatility, extreme market volatility during the smart contract lockup wait (e.g. a monetary policy announcement) could make the trade unprofitable. Bob can mitigate this risk by widening spreads or avoiding swapping altogether during news announcements.
\end{itemize}

\subsubsection{Risk Profile for the Buyer (Alice)}
      Since Alice is required to initiate the swap by creating the smart contract and locking her funds, she incurs non-refundable gas fees and must wait for the Bitcoin transaction to timeout before reclaiming her funds. If Bob, for any reason, fails to complete his side of the swap (a scenario known as \emph{ghosting}), Alice will lose these minimal gas fees. Although these costs are typically low due to efficient contract design and the use of minimal proxies or Layer~2 solutions, they still represent a potential downside for Alice.
\subsubsection{Mitigation Strategies}

\begin{itemize}[label=\(\bullet\)]
    \item \textbf{Collateral and Timeout Mechanisms:} \\
          The protocol can require Bob to post collateral when locking his funds. If Bob ghosts the swap by failing to finalize the transaction within the designated time window, his collateral is forfeited as a penalty. This not only compensates Alice for her incurred gas fees, but also discourages Bob from repeatedly defaulting.
    
    \item \textbf{Reputation Monitoring:} \\
          Alice is encouraged to use off-chain block explorers, such as Etherscan, to monitor Bob’s interactions with the swap contract. Alternatively, a decentralized reputation system (potentially integrated into the swap interface) could provide an easily interpretable record of Bob’s historical performance. The review process of Bob's address's interactions with the smart contract ensures he has a strong track record of completing swaps without ghosting. High reputation scores indicate consistent behavior, thereby reducing Alice’s risk. Additionally, Alice can verify she is indeed interacting with Bob and not an adversary Eve, since Bob's message containing the swap details is signed using the public key associated with his address.
\end{itemize}

\subsection{Third-Party Facilitation and Fee Structure}

In our protocol, a third party (referred to here as the \emph{facilitator}) can provide a value-added service by initiating the swap contract on behalf of the buyer (Alice) and connecting her with the market maker (Bob). This intermediary role not only simplifies the user experience for Alice but also generates revenue for the facilitator through a fee-based model. In practice, the facilitator could be an exchange aggregator or centralized exchange seeking to profit by connecting buyers and sellers. The process is described in detail below.

\subsubsection{Contract Initiation by the Facilitator}
The facilitator receives swap parameters from Alice, including her ETH receiving address and desired swap conditions. Using a factory contract, the facilitator deploys a new swap instance on Ethereum. This instance is configured with:
\begin{itemize}[label=\(\bullet\)]
    \item Alice's beneficiary address,
    \item The predetermined swap parameters (asset amounts, timeouts, etc.),
    \item A public commitment \( C = H(s_a) \) to the unlocking secret.
    \item The facilitator's beneficiary address to take a fee if the swap is successful.
\end{itemize}
Because the facilitator initiates the contract, they incur the gas fees for this deployment, but their service ensures that the process is streamlined for Alice. The facilitator can also amortize the cost of gas fees for initiating the swap if a customer like Alice occasionally ghosts.

\subsubsection{Matching and Swap Execution}
Before the swap instance is created:
\begin{itemize}[label=\(\bullet\)]
    \item The facilitator matches Alice with a market maker (Bob), who pre-signs his Bitcoin redemption transaction using an adaptor signature. Bob's adaptor signature is constructed such that it remains incomplete until an external condition (e.g., confirmation of the ETH lock-up) is met.
    \item Bob sends his pre-signed Bitcoin transaction details (including the commitment \( C \)) to the facilitator, who then communicates this information to Alice.
\end{itemize}

\subsubsection{Fee Deduction Mechanism}
Upon successful completion of the swap, the funds are released as follows:
\begin{itemize}[label=\(\bullet\)]
    \item The Oracle, after verifying the swap conditions and external triggers, issues a signature that enables Bob to redeem the Bitcoin. This process mathematically reveals the secret \( s_a \), which the facilitator can Hash to feed into the smart contract to release the locked ETH to Alice's designated address.
    \item Simultaneously, the contract finalizes the Bitcoin transaction via the adaptor signature mechanism.
    \item A fee is automatically deducted from the overall payout. Let \( A \) denote the total amount transferred in the swap, and let \( \alpha \) (where \( 0 < \alpha < 1 \)) be the fee fraction. Then the facilitator receives:
    \[A_{\text{fee}} = \alpha A,\]
    while the net amount distributed to the parties is:
    \[A_{\text{net}} = (1 - \alpha) A.\]
\end{itemize}
This fee mechanism is embedded in the swap contract’s logic, ensuring that the facilitator is fairly compensated only upon successful swap completion.

\subsubsection{Economic and Operational Benefits}
\begin{itemize}[label=\(\bullet\)]
    \item \textbf{Enhanced Liquidity and Reduced Friction:} \\
          By initiating the contract on Alice's behalf, the facilitator lowers the technical barrier for buyers, enabling more users to participate in decentralized swaps without needing to manage contract deployments or intricate protocols. Popular, easy to use interfaces like Coinbase already support the required Taproot features for our protocol to work for Alice.
    \item \textbf{Low-Risk for Market Makers:} \\
          Since Alice's funds are locked first, Bob (the market maker) faces no risk of being scammed or experiencing delays, allowing him to execute his part of the transaction swiftly. More importantly, a well-funded facilitator with high-speed reliable servers acting on Alice's behalf ensures all parties receive their funds as fast as possible.
    \item \textbf{Revenue Generation:} \\
          The fee structure incentivizes the facilitator to maintain a high-quality service and reliable matching, while also providing a sustainable revenue stream. Although the facilitator may incur gas fees during contract initiation, these costs are offset with fee revenue, and the overall process remains efficient, especially when combined with Layer~2 solutions. It would also let Alice buy ETH without already having Ether to pay gas fees, opening up the protocol to more potential customers.
\end{itemize}

\subsubsection{Flexibility in Oracle and Third-Party Integration}
The authors of this paper believe that the true power of cryptocurrencies lies in their decentralized nature, a return to Satoshi Nakamoto's original vision in creating Bitcoin\cite{nakamoto2008bitcoin}. Unfortunately, many recent developments in the broader cryptocurrency space have been disingenuous in their aims. Nearly two decades after his white paper, the community has strayed further from his goal of decentralization as cryptocurrency has come under scrutiny by regulators, and for good reason. Most contemporary white papers feign technical innovation as a guise for profit-seeking to enrich developers and their investors before an ICO through pre-mines. Thus, it is unsurprising that cryptocurrency in the eyes of regulators and laymen have been synonymous with fraud with the omnipresence of ``memecoins" and ``rugpulls." They remain short-sighted in their vision, unable to believe that fair adoption and profit-seeking behavior are not mutually exclusive. If developers and investors truly believe in the value of their token, they then stand to make a fortune acquiring tokens at launch and holding them for the long-run. The very concept of a pre-mine creates a vicious cycle: skepticism about the coin's value disincentivizes long-term holding, which in turn drives down the price of the very asset investors expect to appreciate in value. The flexibility of the facilitator model we propose allows the unlocking process to be compatible with any decentralized oracle solution, rather than one dictated by the authors. The presence of an established decentralized ecosystem ensures that the process remains secure and not merely a tool to be exploited for centralized control, whether by authoritarian governments or unscrupulous actors.

\section{Conclusion}
This paper introduces an innovative method for executing atomic swaps that leverages pre-signed adaptor signatures built on Schnorr signatures and Bitcoin’s Taproot upgrade. By replacing traditional HTLCs with PTLCs, our approach addresses many of the inherent limitations of earlier methods: long waiting times, rigid on-chain scripting, and privacy concerns. Our design ensures that the buyer initiates the swap via the smart contract and locks funds in the Taproot output. The market maker, in turn, only completes their transaction after the unlocking secret is released by an external oracle. This structure allows the market maker to finalize their side of the trade almost instantaneously once the buyer has committed, enabling high-frequency trading without the need for centralized order books. The implications of our work extend far beyond technical enhancements. By enabling trustless, rapid, and decentralized asset exchanges, our proposal could dramatically increase liquidity in the DeFi ecosystem and empower a broader range of users to swap cryptocurrencies independently. Ultimately, this fosters a more resilient and secure financial system where individuals are not completely reliant on centralized exchanges for cross-chain transactions with decent spreads. We hope that these findings inspire further innovation in cross-chain trading protocols and contribute to a future where more people can safely and efficiently manage their assets in a decentralized manner.

\section{Future Work}
While we have developed a theoretical framework for the mathematical underpinnings of our idea, we intend to release a v1.0 of this white paper with a real-world code implementation, as well as an API for a facilitator to easily conduct these swaps. We also intend to research implementing our protocol using Layer 2 solutions like Bitcoin's Lightning Network or Zk-Rollups in Ethereum. In the future, this could reduce transaction times for market makers from 15 seconds to approximately 2 seconds or less, driving more liquidity and tighter spreads to the DeFi ecosystem.

\bibliographystyle{plainnat}  
\bibliography{references}     

\begin{thebibliography}{9}
\providecommand{\natexlab}[1]{#1}
\providecommand{\url}[1]{\texttt{#1}}
\expandafter\ifx\csname urlstyle\endcsname\relax
  \providecommand{\doi}[1]{doi: #1}\else
  \providecommand{\doi}{doi: \begingroup \urlstyle{rm}\Url}\fi

\bibitem[et~al.(2021)]{Chainlink2.0}
Breidenbach et~al.
\newblock Chainlink 2.0: Next steps in the evolution of decentralized oracle networks.
\newblock Whitepaper, 2021.
\newblock URL \url{https://chain.link/whitepaper-v2}.
\newblock Chainlink Labs; Accessed: March 15, 2025.

\bibitem[Gugger(2020)]{Gugger2020}
Joël Gugger.
\newblock Bitcoin–monero cross-chain atomic swap.
\newblock arXiv preprint, 2020.
\newblock URL \url{https://eprint.iacr.org/2020/1126}.
\newblock Accessed: March 15, 2025.

\bibitem[Hoenisch and Soriano~del Pino(2021)]{Hoenisch2021}
Philipp Hoenisch and Lucas Soriano~del Pino.
\newblock Atomic swaps between bitcoin and monero.
\newblock arXiv preprint, 2021.
\newblock URL \url{https://arxiv.org/abs/2101.12332v2}.
\newblock Accessed: March 14, 2025.

\bibitem[Nakamoto(2008)]{nakamoto2008bitcoin}
Satoshi Nakamoto.
\newblock Bitcoin: A peer-to-peer electronic cash system.
\newblock \url{https://bitcoin.org/bitcoin.pdf}, 2008.
\newblock Accessed: March 14, 2025.

\bibitem[Schnorr(1990)]{Schnorr1990}
Claus-Peter Schnorr.
\newblock Efficient identification and signatures for smart cards.
\newblock In \emph{Advances in Cryptology—CRYPTO '89}, pages 239--252, Berlin, Heidelberg, 1990. Springer-Verlag.
\newblock Universität Frankfurt.

\bibitem[Tu et~al.(2024)Tu, Zhang, and Chen]{Tu2024}
Binbin Tu, Min Zhang, and Yu~Chen.
\newblock Efficient {ECDSA}-based adaptor signature for batched atomic swaps.
\newblock Cryptology {ePrint} Archive, Paper 2024/140, 2024.
\newblock URL \url{https://eprint.iacr.org/2024/140}.

\bibitem[{UnstoppableSwap}(2025)]{UnstoppableSwap_core}
{UnstoppableSwap}.
\newblock Unstoppableswap/core.
\newblock GitHub repository, 2025.
\newblock URL \url{https://github.com/UnstoppableSwap/core}.
\newblock Retrieved from GitHub. Accessed: March 14, 2025.

\bibitem[Wuille et~al.(2020)Wuille, Nick, and Towns]{bip341_github}
Pieter Wuille, Jones Nick, and Anthony Towns.
\newblock Taproot: {SegWit} version 1 spending rules.
\newblock GitHub repository, 2020.
\newblock URL \url{https://github.com/bitcoin/bips/blob/5767f44/bip-0341.mediawiki}.
\newblock Accessed: March 15, 2025.

\bibitem[You et~al.(2024)]{You2024}
Shengewei You et~al.
\newblock A multi-party, multi-blockchain atomic swap protocol with universal adaptor secret.
\newblock arXiv preprint, 2024.
\newblock URL \url{https://arxiv.org/abs/2406.16822}.
\newblock Accessed: March 14, 2025.

\end{thebibliography}

\end{document}